\documentclass[11pt,epsfig]{article} 
\usepackage{layout}

\usepackage[vcentermath,enableskew]{youngtab}
\usepackage{ytableau}
\usepackage{tikz}
\usepackage{enumerate}
\usepackage{float}
\usepackage{subfig}
\usepackage{color}
\usepackage{xcolor}
\usepackage{slashed}
\usepackage{multirow}
\usepackage{cite}

\let\counterwithin\relax
\usepackage{chngcntr}
\usepackage{amssymb, amsmath,mathrsfs}
\usepackage{braket,slashed,bm}

\usepackage{graphics}
\usepackage{graphicx}
\usepackage{epsf}
\usepackage{epsfig}
\usepackage{float}
\usepackage{makecell}
\usepackage{multirow}
\usepackage{color}
\usepackage{xcolor}
\usepackage{simplewick}
\usepackage{textgreek}

\usepackage[utf8]{inputenc}
\usepackage{amsmath}
\usepackage{amssymb}
\usepackage{subfig}
\usepackage[normalem]{ulem}

\usepackage{mathtools}
\usepackage{amsmath}
\usepackage{diagbox}

\usepackage{longtable}
\usepackage{ltablex,booktabs}
\usepackage{tabularray}
\newcolumntype{Y}{>{\centering\arraybackslash}X}
\newcolumntype{C}[1]{>{\centering\arraybackslash}m{#1}}
\usepackage{hhline}
\usepackage{comment}
\usepackage[bb = boondox]{mathalpha}
\usepackage{scrextend}
\usepackage{pifont}
\newcommand{\xmark}{\ding{55}}

\newcommand\undermat[2]{
	\makebox[0.5pt][l]{$\smash{\underbrace{\phantom{%
					\begin{matrix}#2\end{matrix}}}_{ \let\scriptstyle\textstyle\text{\large $#1$}}}$}#2}
\newcommand\overmat[2]{
	\makebox[-1pt][l]{$\smash{\overbrace{\phantom{%
					\begin{matrix}#2\end{matrix}}}^{ \let\scriptstyle\textstyle\text{\large $#1$}}}$}#2}    
\usepackage{tikz}
\usepackage[vcentermath]{youngtab}
\usepackage{slashed} 
\usepackage[font=small]{caption} 
\usepackage{times} 

\usepackage{bm}       
\usepackage{bbm} 

\usepackage{relsize}  

\usepackage{autobreak}  

\usepackage[colorlinks=true,
linkcolor=blue,
urlcolor=red,
citecolor=red]{hyperref}

\usepackage{makeidx} 
\usepackage{bbding} 
\usepackage{listings} 
\usepackage{ytableau}
\usepackage[utf8]{inputenc}
\usepackage[T1]{fontenc}
\usepackage{lmodern}

\usepackage[most]{tcolorbox}
\usepackage{color, colortbl}

\ytableausetup
{boxsize=1.0em}

\long\def\rpl#1!!#2!!{\textcolor{red}{#1} \textcolor{blue}{#2}}

\usepackage[top=1in, left=0.95in, bottom=1.1in, right=0.95in]{geometry}
\def\baselinestretch{1.27}
\usepackage[toc]{appendix}

\newcommand{\beq}{\begin {equation}}
\newcommand{\eeq}{\end   {equation}}
\newcommand{\bea}{\begin {eqnarray}}
\newcommand{\eea}{\end   {eqnarray}}
\newcommand{\beqa}{\begin {eqnarray}}
\newcommand{\eeqa}{\end   {eqnarray}}
\newcommand{\baa}{\begin {array}   }
\newcommand{\eaa}{\end   {array}   }
\newcommand{\bit}{\begin {itemize} }
\newcommand{\eit}{\end   {itemize} }
\newcommand{\be }{\begin {equation}}
\newcommand{\ee }{\end   {equation}}

\newcommand{\sEFT}{\textit{sEFT}}
\newcommand{\phiEFT}{\textit{\straightphi EFT}}

\newcommand{\gEFT}{\textit{GEFT}}
\newcommand{\chiEFT}{\textit{\textchi EFT}}
\newcommand{\mchiEFT}{\textit{m\textchi EFT}}
\newcommand{\dchiEFT}{\textit{d\textchi EFT}}
\newcommand{\VEFT}{\textit{VEFT}}

\newcommand{\xcheckmark}{\checkmark\kern-1.1ex\raisebox{.7ex}{\rotatebox[origin=c]{125}{--}}}

\begin{document}

\begin{center}






 {\Large \textbf  {Complete EFT Operator Bases for Dark Matter and Weakly-Interacting Light Particle}}\\[10mm]

Huayang Song$^{a}$\footnote{huayangs@itp.ac.cn}, Hao Sun$^{a, b}$\footnote{sunhao@itp.ac.cn}, Jiang-Hao Yu$^{a, b, c, d, e}$\footnote{jhyu@itp.ac.cn}\\[10mm]

\noindent 
$^a${\em \small CAS Key Laboratory of Theoretical Physics, Institute of Theoretical Physics, Chinese Academy of Sciences,    \\ Beijing 100190, P. R. China}  \\
$^b${\em \small School of Physical Sciences, University of Chinese Academy of Sciences,   Beijing 100049, P.R. China}   \\
$^c${\em \small Center for High Energy Physics, Peking University, Beijing 100871, China} \\
$^d${\em \small School of Fundamental Physics and Mathematical Sciences, Hangzhou Institute for Advanced Study, UCAS, Hangzhou 310024, China} \\
$^e${\em \small International Centre for Theoretical Physics Asia-Pacific, Beijing/Hangzhou, China}\\[10mm]

\date{\today}   
          
\end{center}

\begin{abstract}

The standard model can be extended to include weakly-interacting light particle (WILP): real or complex singlet scalar, Majorana or Dirac neutral fermion, neutral or hidden-charged vector boson, etc. Imposing the $Z_2$ symmetry, these particles can be lifted as the weakly-interacting massive particle (WIMP), the candidate of dark matter. Instead, imposing the shift symmetry on the scalar components gives rise to the axion-like particle, dark photon, etc. Utilizing these light degrees of freedom along with the standard model particles and imposing different symmetries, we construct the complete and independent sets of effective operators up to dimension eight with the Young tensor technique, consistent with counting from the Hilbert series.   
	

\end{abstract}

\newpage

\setcounter{tocdepth}{3}
\setcounter{secnumdepth}{3}

\tableofcontents

\setcounter{footnote}{0}

\def\baselinestretch{1.5}
\counterwithin{equation}{section}

\newpage

\section{Introduction}
The Standard Model (SM), being one of the most successful theories in particle physics, still leaves a lot of questions unanswered, e.g. matter-antimatter asymmetry, neutrino oscillation, nature of dark matter (DM), which motivates both the theorists and experimentalists to search for new physics (NP) beyond the SM. However, direct searches, especially the Large Hadron Collider (LHC), have not yielded anything signals beyond the Standard Model (BSM). To explain the null results, physicists usually assume that there is a considerable energy gap between the SM and NP particles, which motivates the development of the standard model effective field theory (SMEFT) by parameterizing the impacts of the BSM physics into the Wilson coefficients of the higher dimensional operators in a model-independent way. 

Alternatively, new particles can be light, varying from tens of GeV down to a few keV~\cite{Beacham:2019nyx,Prebys:2022qig}, yet have very weak couplings to the SM particles, hereafter denoted as a weakly-interacting light particle (WILP)~\footnote{The terminology WILP is the analog~\cite{Feng:2019bci} of weakly-interacting massive particle (WIMP), in which a $Z_2$ symmetry is typically further imposed. Similar names includes weakly-interacting slim particle (WISP)~\cite{Ringwald:2012hr}, light weakly coupled particle (WCP)~\cite{Essig:2013lka}, and feebly interacting particle (FIP)~\cite{Lanfranchi:2020crw}, addressing on axion and axion-like particle, light dark matter, and portal dark matter, etc. Here WILP is used to describe all the above types of particles. }. Though in this case, their interaction with the SM can be governed by renormalizable Lagrangian, the signatures at the LHC are trigger-limited due to limited energy resolution or relatively long lifetimes induced by small couplings. To naturally suppress the couplings, new degrees of freedom (DOFs) are ``sterile'' (neutral) under the SM gauge interactions, or in other words, are SM gauge singlets. Although singlet under the SM gauge group, these particles could still interact with some ``dark sector'' or be charged under hidden symmetry. This scenario is very common in many existing models featuring baryogenesis, sterile neutrino, and dark matter. The minimal models classified via the particle's spin are well studied in case of scalar~\cite{Espinosa:1993bs, OConnell:2006rsp, Barger:2006sk, Ahriche:2007jp, Profumo:2007wc, Barger:2007im, Barger:2008jx, Gonderinger:2009jp, Espinosa:2011ax, Pruna:2013bma, Chen:2014ask, Gorbahn:2015gxa, Dawson:2015haa, Costa:2015llh, Robens:2016xkb, Patt:2006fw}, fermion (heavy neutral lepton, or HNL)~\cite{Minkowski:1977sc, Yanagida:1979as, Gell-Mann:1979vob, Mohapatra:1979ia} and vector (dark photon)~\cite{Okun:1982xi, Galison:1983pa, Holdom:1985ag} with only a few leading order operators of dimension 4 or lower. Significant progress in searching for such kind of particles and testing these models can be achieved by the undergoing or proposed intensity-frontier experiments at the LHC, e.g. ANUBIS~\cite{Bauer:2019vqk, Hirsch:2020klk, Dreiner:2020qbi}, CODEX-b~\cite{Gligorov:2017nwh, Dey:2019vyo, Aielli:2019ivi, Aielli:2022awh}, FASER~\cite{Feng:2017uoz, FASER:2018ceo, FASER:2018bac, FASER:2022hcn, FASER:2021ljd,FASER:2021cpr}, MATHUSLA~\cite{Chou:2016lxi, Curtin:2017izq, Evans:2017lvd, Curtin:2018mvb, Curtin:2018ees, MATHUSLA:2018bqv,  MATHUSLA:2019qpy, Alimena:2019zri, Alidra:2020thg, MATHUSLA:2020uve, MATHUSLA:2022sze, Bose:2022obr}, fixed-target experiments such as NA62/64~\cite{Lanfranchi:2017wzl, NA62:2017rwk, NA64:2016oww, Gninenko:2019qiv, Banerjee:2019pds, NA64:2020qwq, Gninenko:2300189, Gninenko:2640930}, Sea/Spin/Darkquest~\cite{SeaQuest:2017kjt, Liu:2017ryd, Berlin:2018pwi, Batell:2020vqn, Blinov:2021say, Apyan:2022tsd, Tsai:2019buq}, or SHiP~\cite{Bonivento:2013jag, Alekhin:2015byh, SHiP:2015vad}, and new search strategies at the LHC main dectectors ATLAS and CMS~\cite{Gershtein:2017tsv, Liu:2018wte, Lee:2018pag, Alimena:2019zri, Liu:2019ayx, Liu:2020vur, Gershtein:2020mwi, Chiu:2021sgs, Fischer:2021sqw, Bose:2022obr, ATLAS:2022pib, ATLAS:2022cob, CMS:2016ybj, CMS:2015pca, CMS:2022qej, ATLAS:2018rjc, CMS:2017kku, ATLAS:2022zhj, CMS:2020iwv, ATLAS:2021mdj, CMS:2019zxa, CMS:2019qjk, ATLAS:2019qrr, CMS:2021sch, ATLAS:2022vyq, ATLAS:2022gvp}. Therefore a more general model-independent framework for phenomenology studies on WILPs is demanded from both theoretical and experimental perspectives.


Though a single singlet with only renormalizable interactions to the SM, like those minimal models, can be a reasonable theory, extra states generally exist in the ultraviolet (UV) completions. In the case that such DOFs are too heavy to be observed in current and near-future experiments, integrating them out could result in higher dimensional operators describing interactions between light singlets and SM particles. The simplest realization of such a scenario is hidden valley models~\cite{Strassler:2006im} where a hidden sector (WILPs) only communicates with the visible sector via a heavy mediator. Even without suppressed renormalizable couplings, such interactions are naturally suppressed by powers of some heavy particles' mass scale and make the singlets possible WILP candidates. An effective field theory (EFT) description including higher dimensional operators beyond the renormalizable ones just provides us with the technique for WILPs studies. The well-known model for axion or axion-like particles (ALPs)~\cite{Peccei:1977hh, Peccei:1977ur, Weinberg:1977ma, Wilczek:1977pj} just falls into this category since the leading order interactions only show up at dimension 5. Following this direction, we have already presented the EFTs for the axion, ALP and dark photon in Ref.~\cite{Song:2023lxf} up to dimension 8. As also shown in that paper, a certain class of EFTs, e.g. EFT for majoron, only give observable phenomena going beyond the lower dimensional operators ($dim$-8 for majoron). However, that paper only presents EFTs for a real Goldstone and a dark photon without considering more general particle species, e.g. real/complex scalar, real/complex vector. Hence we present the complete and independent operators involving singlet with arbitrary spins ($\leq 1$) up to dimension 8 in this paper. Such complete bases are useful if the Wilson coefficients of the leading operators accidentally vanish due to some unknown UV physics and could also lead to non-trivial phenomenology in the experimental searches of these BSM particles. Imposing further symmetries, our general EFTs can be reduced to describe some specific models. For example, the singlet can serve as a DM candidate, the WIMP particle, if one requires the singlet to possess a $\mathbf{Z}_2$ symmetry, and EFT for real scalar with a shift symmetry describes an axion or ALP.

The singlet extension respecting a $\mathbf{Z}_2$ symmetry, under which the singlet is odd and stable, is also known as Dark Matter EFT (DMEFT), or Dark SMEFT (DSMEFT). Some early works along this direction usually assume that the singlet is a WIMP and focus on some specific operators relating to DM generation, (in)direct detection or DM collider searches~\cite{Beltran:2008xg, Cirelli:2008pk, Shepherd:2009sa, Cao:2009uw, Beltran:2010ww, Agrawal:2010fh, Fitzpatrick:2010em, Goodman:2010yf, Goodman:2010ku, Goodman:2010qn, Zheng:2010js, Rajaraman:2011wf, Yu:2011by, March-Russell:2012elz, Carpenter:2012rg}. Since then operators up to dimension eight for particular singlet spins are described in several references. For instance, a complete set of operators of dimension $\leq 6$ for SMEFT extension with a complex scalar DM field in a singlet (also doublet or triplet) representation is presented in Ref.~\cite{DelNobile:2011uf}. In Ref.~\cite{Duch:2014xda}, a basis of operators of dimension $\leq 6$ for an EFT with the SM and $\mathbf{Z}_2$ DM particles have been given, assuming the non-SM particles with spin$\leq 1$. A basis of effective operators up to dimension 7 has been given in Ref.~\cite{Brod:2017bsw} for scalar and fermionic DM in general SM representations. Ref.~\cite{Criado:2021trs} extends the discussion to also include spin 1 DM particles in a general non-redundant basis up to dimension 6 assuming that DM particles are odd under a $\mathbf{Z}_2$ symmetry.
However, as the authors pointed out~\cite{Criado:2021trs}, they do not impose any restrictions 
from the extra gauge symmetries of the added spin 1 particle, which might generally lead to difficulties to UV complete the operators. Further, they do not include any self-interaction operators of the DM particles.
Ref.~\cite{Barducci:2021egn} also presents parts of $dim$-6 operators (dipole operators and operators generating kinetic terms or kinetic mixing) describing interactions between spin-1 dark photon and the SM, and discusses their possible UV completions and phenomenological implications. Recently a general DSMEFT including scalar, fermion and vector DM fields without imposing any underlying symmetry on the DM fields are constructed in Ref.~\cite{Aebischer:2022wnl}. However, since they only combine gauge invariant operators (building from field strength tensors) with mass terms of vector particles, some operators discussed in Refs.~\cite{Lebedev:2011iq, Duch:2014xda, Reece:2018zvv, Kribs:2022gri, Song:2023lxf} are missed. 
Without assuming that the additional particles serve as DM (roughly speaking there is no $\mathbf{Z}_2$ symmetry to stabilize the particles), a study of portal effective theories is performed in Ref.~\cite{Arina:2021nqi}, in which all portal operators at the electroweak scale up to dimension five are constructed. Complete bases of operators including the SM fields together with sterile right-handed neutrinos are obtained in Refs.~\cite{delAguila:2008ir, Aparici:2009fh} for $dim$-5, Refs.~\cite{delAguila:2008ir, Liao:2016qyd, Bischer:2019ttk, Bischer:2020sop} for $dim$-6, Refs.~\cite{Bhattacharya:2015vja, Liao:2016qyd} for $dim$-7, and Ref.~\cite{Li:2021tsq} up to $dim$-9. However, to the best of the authors' knowledge, beyond dimension five, the complete and non-redundant operator bases of singlet SMEFT extensions (\sEFT s) including scalar (\phiEFT), fermion (\chiEFT) or vector (\VEFT) singlet are not known. In this paper, we construct and enumerate these complete and independent operator bases in a gauge-invariant way with the help of the Young tensor technique and also check the numbers of independent operators with the Hilbert series counting method~\cite{Benvenuti:2006qr, Feng:2007ur, Lehman:2015via, Henning:2015alf, Lehman:2015coa, Henning:2017fpj, Graf:2020yxt, Graf:2022rco, Sun:2022aag}. We hope our work closes the gap and presents useful operators bases for further phenomenological research.

The paper is organized as follows. In Sec.~\ref{sec:opsbbs&YTT} we briefly review the building blocks and the Young tensor method to construct the complete and independent effective operator. We then discuss the general scalar extensions (\phiEFT s) in Sec.~\ref{sec:operator4scalar} and list the complete operators involving real or complex scalar singlets up to dimension 8. The EFT describing a Goldstone, axion or ALP and majoron can be further obtained by imposing a shift symmetry. In Sec.~\ref{sec:operator4fermion}, both Majorana and Dirac fermion singlet extensions (\chiEFT s) are described. The effect of a global $SO(2)$ symmetry in the Dirac case is discussed. Operators involving either type of fermion are listed. Sec.~\ref{sec:operator4photon} lists the complete operators up to $dim$-8 of real and complex vector singlet extensions of the SMEFT (\VEFT s) by introducing Goldstone fields as the longitudinal modes of the vector fields. We reach our conclusion in Sec.~\ref{sec:conclu}.

\section{The Strategy of the Basis Construction}
\label{sec:opsbbs&YTT}

The Young tensor method is a systematic way to construct independent and complete higher-dimension effective operators, which has been used in the SMEFT~\cite{Li:2020gnx,Li:2020xlh}, the LEFT~\cite{ Li:2020tsi} and HEFT~\cite{Sun:2022ssa,Sun:2022snw}. In this section, we would briefly review the main point of the method, present the building blocks, and declare the notations used in the paper.

\subsection{Amplitude-Operator correspondence}

A connected amplitude is obtained by the interaction sandwiched between the in- and out-states,
\begin{equation}
\label{eq:cor}
\mathcal{A} = \langle f|\mathcal{O}|i\rangle\,,
\end{equation}
where $\mathcal{O}$ is the operator implemented for the interaction, and $|i/f\rangle$ is the in-/out-state. Actually, Eq.~\ref{eq:cor} relates the operator $\mathcal{O}$ to an on-shell amplitude, which is the foundation of the Young tensor method.

In particular, such amplitude-operator correspondence makes it possible to translate various fields in the effective field theories to Weyl spinors in the amplitude, which can be presented in terms of their helicities $h$ as
\begin{align}
    h=0: & \quad\phi\sim 1\notag\\
    h=-\frac{1}{2}: &\quad \psi\sim \lambda_\alpha\notag\\
    h=\frac{1}{2}: &\quad \psi^\dagger\sim \Tilde{\lambda}^{\dot{\beta}} \notag \\
    h=-1: &\quad X_L\sim \lambda_\alpha\lambda_\beta \notag \\
    h=1: &\quad X_R\sim \tilde{\lambda}^{\dot{\alpha}}\tilde{\lambda}^{\dot{\beta}} \,.\label{eq:AOC}
\end{align}
The left-handed fermion $\psi$ is related to the left-handed spinor $\lambda_\alpha$, while the right-handed fermion $\psi^\dagger$ is related to the right-handed spinor $\Tilde{\lambda}^{\dot{\beta}}$. The left-/right-handed field-strength tensor is identified with the symmetric product of left-/right-handed spinors. In addition, the on-shell momentum is referred to the product of left- and right-handed spinors 
\begin{equation}
p\sim \lambda_\alpha\tilde{\lambda}^{\dot{\beta}}\,.
\end{equation}

According to the correspondences in Eq.~\ref{eq:AOC}, we can make the amplitude-operator correspondence more explicit,
\begin{equation}
\mathcal{O}\sim \prod^n \langle ij\rangle\prod^{\Tilde{n}} [kl]\,,\label{eq:cor2}
\end{equation}
where we have used the spinor contraction brackets
\begin{align}
    \langle ij\rangle &= \lambda_{i}^\alpha \lambda_{j\alpha} = -\lambda_{j}^\alpha \lambda_{i\alpha} = -\langle ji\rangle\,, \label{eq:contr1}\\
    [ij] &= \tilde{\lambda}^i_{\dot{\alpha}}\tilde{\lambda}^{j\dot{\alpha}} = -\tilde{\lambda}^i_{\dot{\alpha}}\tilde{\lambda}^{j\dot{\alpha}} = -[ji]\,.\label{eq:contr2}
\end{align}

In summary, we can interpret all the fields in the SM together with the various singlet extensions to spinor forms. Restoring the SM gauge symmetry $SU(3)_C\times SU(2)_W\times U(1)_Y$ and specific added abelian symmetry $U(1)$, we list all the building blocks in Tab.~\ref{tab:buildingblocks}, where the charges $q_S\,,q_u\,,q_M\,,q_D\,,q_V$ under the added $U(1)$ symmetry are assumed.

\begin{table}[]
    \centering
 \begin{tabular}{c|c|c|c|c|c||c}
\hline
Sector & Building block & Lorentz group & $SU(3)_C$ & $SU(2)_W$ & $U(1)_Y$ & $U(1)$\\
\hline
\multirow{9}{*}{SM} & $G_{L/R}$ & $(1,0)/(0,1)$ & $\mathbf{8}$ & $\mathbf{1}$ & 0 & 0 \\ 
\cline{2-7}
& $W_{L/R}$ & $(1,0)/(0,1)$ & $\mathbf{1}$ & $\mathbf{3}$ & 0 & 0 \\
\cline{2-7}
& $B_{L/R}$ & $(1,0)/(0,1)$ & $\mathbf{1}$ & $\mathbf{1}$ & 0 & 0 \\
\cline{2-7}
& $L/L^\dagger$ & $(\frac{1}{2},0)/(0,\frac{1}{2})$ & $\mathbf{1}$ & $\mathbf{2}$ & $\mp \frac{1}{2}$ & 0 \\
\cline{2-7}
& $e_c/e_c^\dagger$ & $(\frac{1}{2},0)/(0,\frac{1}{2})$ & $\mathbf{1}$ & $\mathbf{1}$ & $\pm 1$ & 0 \\
\cline{2-7}
& $Q/Q^\dagger$ & $(\frac{1}{2},0)/(0,\frac{1}{2})$ & $\mathbf{3}/\overline{\mathbf{3}}$ & $\mathbf{2}$ & $\pm \frac{1}{6}$ & 0 \\
\cline{2-7}
& $u_c/u_c^\dagger$ & $(\frac{1}{2},0)/(0,\frac{1}{2})$ & $\overline{\mathbf{3}}/\mathbf{3}$ & $\mathbf{1}$ & $\mp \frac{2}{3}$ & 0 \\
\cline{2-7}
& $d_c/d_c^\dagger$ & $(\frac{1}{2},0)/(0,\frac{1}{2})$ & $\overline{\mathbf{3}}/\mathbf{3}$ & $\mathbf{1}$ & $\pm \frac{1}{3}$ & 0 \\
\cline{2-7}
& $H/H^\dagger$ & $(0,0)$ & $\mathbf{1}$ & $\mathbf{2}$ & $\pm \frac{1}{2}$ & 0 \\
\hline
\hline
Real scalar & s & $(0,0)$ & $\mathbf{1}$ & $\mathbf{1}$ & 0 & 0 \\
\hline
\hline
Complex Scalar & $S/S^\dagger$ & $(0,0)$ & $\mathbf{1}$ & $\mathbf{1}$ & 0 & $\pm q_S$ \\
\hline
\hline
Real Goldstone & $u_\mu = D_\mu s$ & $(\frac{1}{2},\frac{1}{2})$ & $\mathbf{1}$ & $\mathbf{1}$ & 0 & 0 \\
\hline
\hline
\multirow{2}{*}{Complex Goldstone} & $u_\mu=D_\mu S$ & \multirow{2}{*}{$(\frac{1}{2},\frac{1}{2})$} & \multirow{2}{*}{$\mathbf{1}$} & \multirow{2}{*}{$\mathbf{1}$} & \multirow{2}{*}{0} & \multirow{2}{*}{$\pm q_u$} \\
& $u^\dagger_\mu= D_\mu S^\dagger$ & & & & \\
\hline
\hline
Majorana fermion & $\chi/\chi^\dagger$ & $(\frac{1}{2},0)/(0,\frac{1}{2})$ & $\mathbf{1}$ & $\mathbf{1}$ & 0 & $\pm q_M$ \\
\hline
\hline
\multirow{2}{*}{Dirac fermion} & $\chi_L/\chi^\dagger_L$ & $(\frac{1}{2},0)/(0,\frac{1}{2})$ & $\mathbf{1}$ & $\mathbf{1}$ & 0 & $\pm q_D$ \\
\cline{2-7}
& $\chi_{Rc}/\chi_{Rc}^\dagger$ & $(\frac{1}{2},0)/(0,\frac{1}{2})$ & $\mathbf{1}$ & $\mathbf{1}$ & 0 & $\mp q_D$ \\
\hline
\hline
Real Vector & $X_L/X_R$ & $(1,0)/(0,1)$ & $\mathbf{1}$ & $\mathbf{1}$ & 0 & 0 \\
\hline
\hline
\multirow{2}{*}{Complex Vector} & $X_L/X_R$ & $(1,0)/(0,0)$ & $\mathbf{1}$ & $\mathbf{1}$ & 0 & $+q_V$ \\
\cline{2-7}
& $X_L^\dagger/X_R^\dagger$ & $(1,0)/(0,1)$ & $\mathbf{1}$ & $\mathbf{1}$ & 0 & $-q_V$ \\
\hline
\hline
\end{tabular}
    \caption{The extension of the SM, where the irreducible representations of the Lorentz group are written in terms of the decomposition $SO(1,3)\sim SU(2)_l\times SU(2)_r$, and $q_S\,,q_u\,,q_M\,,q_D\,,q_V$ are the possible charges of the additional complex particles. To identify the complex Goldstone as the complex vector's longitudinal mode, we further require $q_u=q_V$.}
    \label{tab:buildingblocks}
\end{table}

\subsection{Young Tensor Technique}
Though the amplitude-operator correspondence can translate any effective operators to on-shell amplitudes, it is usually the case where we need to utilize the spinors to construct the corresponding operators and this can be done systematically by the Young tensor method~\cite{Henning:2019mcv, Henning:2019enq,Li:2020gnx,Li:2020xlh,Li:2022tec}.

As shown in Eq.~\ref{eq:cor2}, $n$ and $\Tilde{n}$ are the numbers of angle brackets $\langle .\rangle $ and squared brackets $[.]$ respectively, which can be determined by the fields and the numbers of derivatives in the effective operator. Even given certain values of $n\,,\Tilde{n}$, there are still many algebra relations of spinors making the construction complicated. For example, the Schouten identities of spinors are related to the Fierz identities of the fermion fields, and the momentum conservation is equivalent to the total derivative of operators, both of which should be taken care of to avoid the redundancy. Actually, it has been argued~\cite{Henning:2019enq,Henning:2019mcv,Li:2020gnx} that for an operator type with $N$ fields and $n_D$ derivatives, all such redundancies can be eliminated by focusing on the Young diagram of the shape
\begin{equation}
    \begin{tikzpicture}
\filldraw [draw = black, fill = cyan] (10pt,10pt) rectangle (22pt,22pt);
\filldraw [draw = black, fill = cyan] (10pt,22pt) rectangle (22pt,34pt);
\draw [densely dotted] (24pt,22pt) -- (32pt,22pt);
\filldraw [draw = black, fill = cyan] (10pt,46pt) rectangle (22pt,58pt);
\filldraw [draw = black, fill = cyan] (10pt,58pt) rectangle (22pt,70pt);
\draw [densely dotted] (24pt,58pt) -- (32pt,58pt);
\draw [densely dotted] (16pt,36pt) -- (16pt,46pt);
\filldraw [draw = black, fill = cyan] (34pt,10pt) rectangle (46pt,22pt);
\filldraw [draw = black, fill = cyan] (34pt,22pt) rectangle (46pt,34pt);
\filldraw [draw = black, fill = cyan] (34pt,46pt) rectangle (46pt,58pt);
\filldraw [draw = black, fill = cyan] (34pt,58pt) rectangle (46pt,70pt);
\draw [densely dotted] (40pt,36pt) -- (40pt,46pt);
\draw [|<-] (0pt,10pt)--(0pt,34pt);
\draw [|<-] (0pt,70pt)--(0pt,46pt);
\node (n2) at (-5pt,40pt) {\small $N-2$};
x
\draw [|<-] (10pt,80pt)--(22pt,80pt);
\draw [|<-] (46pt,80pt)--(34pt,80pt);
\node (nt) at (28pt,80pt) {\small $\tilde{n}$};

\draw (46pt,58pt) rectangle (58pt,70pt);
\draw (46pt,46pt) rectangle (58pt,58pt);
\draw [densely dotted] (59pt,58pt) -- (67pt,58pt);
\draw (68pt,58pt) rectangle (80pt,70pt);
\draw (68pt,46pt) rectangle (80pt,58pt);
\draw [|<-] (46pt,36pt)--(58pt,36pt);
\draw [|<-] (80pt,36pt)--(68pt,36pt);
\node (nt) at (63pt,36pt) {\small $n$};
    \end{tikzpicture}\,,
\end{equation}
which is called primary Young diagram.

According to group theory, a set of basis can be generated by filling the Young diagram to form the so-called semi-standard Young tableaux (SSYT). In the cases of definite field number $N$ and derivative number $n_D$, the numbers $i$ to be filled range from 1 to $N$, and their repetitions, denoted by $\#i$, are also determined by,   
\begin{equation}
\#i = \frac{1}{2}n_D+\sum_{h_i>0}|h_i|-2h_i,\quad i=1,2,\dots N\,,
\end{equation}
where $(h_1,h_2,\dots,h_N)$ are the helicities. Once the SSYTs are obtained, we can translate them to the amplitudes according to the following rules, 
\begin{equation}
\begin{ytableau}
*(cyan) i_1  \\
*(cyan) i_2  \\
\none[\vdots] \\
*(cyan) i_{N-2} \\
\end{ytableau} \rightarrow \epsilon^{i_1i_2\dots i_{N-2}kl}[kl]\,,\quad
\begin{ytableau}
i  \\
j  \\
\end{ytableau} \rightarrow \langle ij \rangle\,,
\end{equation}
The basis obtained in this way is called the y-basis (Young tableau basis).

Since the effective operators are both Lorentz and gauge invariant, we need to construct independent gauge tensors as well, which is similar to the procedure in the Lorentz sector. For a general gauge group $SU(N)$, the invariant tensors correspond to the so-called singlet Young diagram 
\begin{equation}
\label{eq:singletYD}
\begin{tikzpicture}
\draw (20pt,20pt) rectangle (34pt,34pt);
\draw (34pt,20pt) rectangle (48pt,34pt);
\draw [loosely dotted] (50pt,27pt)--(60pt,27pt);
\draw (62pt,20pt) rectangle (76pt,34pt);
\draw [loosely dotted] (27pt,36pt)--(27pt,46pt);
\draw [loosely dotted] (69pt,36pt)--(69pt,46pt);
\draw (20pt,48pt) rectangle (34pt,62pt);
\draw (34pt,48pt) rectangle (48pt,62pt);
\draw [loosely dotted] (50pt,62pt)--(60pt,62pt);
\draw (62pt,48pt) rectangle (76pt,62pt);
\draw (20pt,62pt) rectangle (34pt,76pt);
\draw (34pt,62pt) rectangle (48pt,76pt);
\draw (62pt,62pt) rectangle (76pt,76pt);
\draw [|<-] (13pt,20pt)--(13pt,41pt);
\draw [|<-] (13pt,76pt)--(13pt,55pt);
\node (N) at (13pt,48pt) {{\small$N$}};
\end{tikzpicture}
\end{equation}
and a set of basis can be generated by the SSYTs of it. According to group theory, all the irreducible representations of $SU(N)$ correspond to Young tableaux isomorphically, thus the construction of invariant tensors is the outer product of the Young tableaux to form the singlet Young tableaux, which is governed by the Littlewood-Richardson rule, and one can translate them in the following way, 
\begin{equation}
\begin{ytableau}
i_1  \\
i_2  \\
\none[\vdots] \\
i_{N} \\
\end{ytableau} \rightarrow \epsilon^{i_1i_2\dots i_{N}}\,,
\end{equation}

The complete operator basis is obtained by the tensor product of the Lorentz and gauge sectors. Nevertheless, if there are repeated fields in the operator type, the flavor structure may lead to further complexity, and some additional redundancies may arise. Thus it is convenient to further transform the above resultant basis to a set that is classified into the subspaces with specific permutation symmetries of the repeated fields, which is called the p-basis. In the Young tensor method, such transformation is completed by idempotent,
\begin{equation}
\mathcal{Y}^{[\lambda]}\,,
\end{equation}
where $[\lambda]$ is the Young diagram representing flavor symmetry. The operators with specific flavor symmetry form the f-basis, which is the final result of this paper.

\subsection{Goldstone Bosons}\label{sec:Adler}
The Goldstone bosons need an extra process to deal with since they are scalars whose amplitudes satisfy the Adler's zero condition~\cite{Adler:1964um, Adler:1965ga, Low:2014nga, Low:2014oga, Cheung:2014dqa, Cheung:2015ota, Low:2019ynd, Dai:2020cpk, Sun:2022ssa, Low:2022iim, Sun:2022snw}. In the soft momentum limit, an operator involving the Goldstones corresponds to an on-shell amplitude satisfying
\begin{align}
    \mathcal{M}(p_\pi)\rightarrow p_\pi\quad\text{for}\quad p_\pi\rightarrow 0.
\end{align}
where we use $p_\pi$ to represent the momentum of an external Goldstone. In Refs.~\cite{Sun:2022ssa, Sun:2022snw} and our previous paper~\cite{Song:2023lxf}, the method to obtain the general Lorentz basis of Goldstone is discussed by reducing scalars amplitudes with imposing the Adler's zero condition. The readers can find detailed procedures and further useful information in those references. The complete and independent operator basis for a Goldstone singlet up to dimension eight is presented in Ref.~\cite{Song:2023lxf} for the first time with this method.


\let\clearpage\relax
\section{Scalar EFT (\phiEFT) Operator Bases\label{sec:operator4scalar}}
Due to the less understanding on the scalar sector of the SM, it is natural to consider the extensions of the SM with only scalars, among which the scalar singlet extension is the simplest and well-studied~\cite{Espinosa:1993bs, OConnell:2006rsp, Barger:2006sk, Ahriche:2007jp, Profumo:2007wc, Barger:2007im, Barger:2008jx, Gonderinger:2009jp, Espinosa:2011ax, Pruna:2013bma, Chen:2014ask, Gorbahn:2015gxa, Dawson:2015haa, Costa:2015llh, Robens:2016xkb}. Since the scalar singlet transforms trivially under gauge group of the SM, it couples to the SM fields only via the Higgs doublet at the renormalizable level, which is often referred to as Dark Higgs model~\cite{Patt:2006fw}.

\label{hyref:real}In the case of a real scalar, the most general renormalizable Lagrangian is given by
\begin{align}
    \mathcal{L}\supset&\frac{1}{2}\partial_\mu s\partial^\mu s-\frac{1}{2}m_s^2 s^2-\frac{\lambda_4}{4}s^4-\frac{\delta_2}{2}H^\dagger Hs^2 \\
    &-\lambda_1 s-\frac{\lambda_3}{3}s^3-\frac{\delta_2}{2}H^\dagger Hs
\end{align}
where the first line preserves a $\mathbb{Z}_2$ symmetry while the terms in the second line explicitly break the $\mathbb{Z}_2$ symmetry. The $\mathbb{Z}_2$ symmetry can guarantee stability of the field $s$ as long as no VEV can be developed. Therefore the $\mathbb{Z}_2$ symmetric version describes the Higgs portal scalar dark matter~\cite{Silveira:1985rk, Burgess:2000yq, Davoudiasl:2004be, Gonderinger:2009jp, Cirelli:2009uv, Biswas:2011td, Cline:2013gha, Arcadi:2017kky, Arcadi:2019lka, Kanemura:2010sh, Alexander:2016aln, Battaglieri:2017aum}. In the presence of a singlet VEV, it can help generate a strong first-order electroweak phase transition (EWPT) as required by Electroweak Baryogenesis (EB)~\cite{Espinosa:1993bs, Ahriche:2007jp, Profumo:2007wc, Barger:2007im, Espinosa:2011ax, Barger:2011vm, Profumo:2014opa, Cline:2012hg}.

\label{hyref:complx}The complex scalar can be treated as two real scalar singlets, corresponding to the real and imaginary parts. However, the complex scalar has an extra $U(1)$ symmetry which gives further relationship among parameters, therefore it is convenient to work directly with a complex scalar field. The most general renormalizable Lagrangian of complex scalar extension invariant under the $U(1)$ symmetry is
\begin{align}
    \mathcal{L}\supset&\partial_\mu S^\dagger \partial^\mu S-m_S^2 S^\dagger S-\frac{\lambda_4}{4}\left(S^\dagger S\right)^2-\frac{\delta_2}{2}H^\dagger H S^\dagger S
\end{align}
Note that in some parameter regions, field $S$ could develop a vev spontaneously breaking symmetry the $U(1)$ symmetry, which results to a massless Goldstone boson. Such a massless degree of freedom that is not phenomenologically viable, and can be avoided by introducing soft breaking terms or explicit breaking terms as in Ref.~\cite{Barger:2008jx, Gonderinger:2012rd}. Including these terms, however, we easily lose control and the number of independent operators increases significantly. In the following, we will only list the operators which possess a $U(1)$ symmetry~\footnote{For the $U(1)$ spontaneously breaking case, one can use a real scalar singlet + a Goldstone singlet, which is beyond our scope --- single field extension, to describe.}. If this $U(1)$ symmetry is not broken spontaneously, the conservation of the $U(1)$ charge stabilizes the scalar and this theory also describes a complex scalar dark matter interacting with the SM particle through the Higgs portal~\cite{Silveira:1985rk, McDonald:1993ex, Burgess:2000yq, Patt:2006fw, Arcadi:2017kky, Arcadi:2019lka}. Though the vanishing VEV complex scalar extension can trigger a strong first-order EWPT, the upper bound on the mass of the Higgs boson is inconsistent with the observed 125 GeV Higgs partilce~\cite{Espinosa:1993bs}. Provided that the scalar has a VEV, this model can contain both the ingredients for a strong first-order EWPT and a dark matter candidate (the pseudo-Goldstone $A$ from the spontaneous symmetry breaking of $U(1)$)~\cite{Barger:2008jx, Cline:2013gha}.

Among the scalars, Goldstone bosons belong to a special class with non-trivial properties. Such particles emerge from spontaneous breaking of some global symmetries~\cite{Nambu:1960tm, Goldstone:1961eq, Goldstone:1962es}, which can be either fundamental or composite. Axion emerging from the spontaneously broken global Peccei-Quinn symmetry $U(1)_\text{PQ}$ to solve the strong CP problem~\cite{Peccei:1977hh, Peccei:1977ur, Weinberg:1977ma, Wilczek:1977pj}, axion-like particles (ALPs) from theories with extra dimensions and string theory models~\cite{Svrcek:2006yi, Arvanitaki:2009fg, Cicoli:2012sz}, and majoron generated in seesaw model due to the lepton number symmetry breaking~\cite{Gelmini:1980re} can all be embedded into an EFT (\gEFT) describing the interactions between the SM and a Goldstone singlet, which also provide solutions to dark matter and so on~\cite{Coleman:1969sm, Callan:1969sn, Marsh:2015xka}. CCWZ formalism serves a general method to construct such effective Lagrangians by requiring a particular symmetry breaking pattern $G/H$. From a bottom-up perspective, the Goldstone is nothing but a scalar enjoying a shift symmetry, whose amplitudes satisfy the Adler’s zero condition as discussed in Sec.~\ref{sec:Adler}. In Ref.~\cite{Song:2023lxf}, we have obtained a complete and independent operator basis for \gEFT~via reducing the basis for a scalar singlet by imposing Adler's zero condition.

In Tab.~\ref{tab:phiEFT}, we present the complete non-redundant operator basis up to $dim$-8 for scalar singlet extension of the SMEFT (\phiEFT), real scalar case (\textbf{Real} \phiEFT), complex scalar case (\textbf{Complex} \phiEFT) and $\mathbf{Z_2}$ dark matter case ($\mathbf{Z_2}$). For completeness, we also mark the operators contributing to axion (\textbf{A}) and majoron (\textbf{M}).

\setlength\LTleft{-0.65in}
{ \small 
\begin{longtable}{@{}C{0.105\textwidth}C{0.145\textwidth}C{0.315\textwidth}C{0.295\textwidth}C{0.085\textwidth}ccc@{}}
\caption{\textcolor{black}{{Complete and independent operators for real (\textbf{Real}), complex scalar (\textbf{Complex}), ALP or Goldstone (\textbf{A}), Majoron (\textbf{M}), $\mathbf{Z}_2$ dark matter ($\mathbf{Z}_2$) classified in canonical dimension. Column \textbf{F} shows the flavor structure in Young tableau if repeated fields (fermions) in operators lead to ambiguity. Operators marked with $\xcheckmark$ for axion in $dim$-5 do not all contribute to an axion EFT, and further transformation is needed to arrive at a non-redundant basis.}}\label{tab:phiEFT}}\\
\hhline{========}
\multicolumn{8}{c}{\textcolor{blue}{{\textbf{\textit{dim}-5 operators}}}}\\
\hhline{========}
{\textbf{Class}} & {\textbf{Type}} & {\textbf{Real}} & {\textbf{Complex}} & {\textbf{F}} & {\textbf{A}} & {\textbf{M}} & {$\mathbf{Z}_2$} \\
\hline\hline
\multirow{3}{*}{
} & $B_L^2 s$ & $s{{B_L}_{\mu\nu}}{{B_L}^{\mu\nu}}$ &  &  & $\xcheckmark$ &  &  \\ \cline{2-8}
\multirow{1}{*}{$F_L^2\phi$} & {$W_L^2 s$} & $s{{{W_L}^{I}}_{\mu\nu}}{{{W_L}^{I}}^{\mu\nu}}$ &  &  & $\xcheckmark$ &  &  \\ \cline{2-8}
 & $G_L^2 s$ & $s{{{G_L}^{A}}_{\mu\nu}}{{{G_L}^{A}}^{\mu\nu}}$ &  &  & $\xcheckmark$ &  &  \\ \hline
\multirow{3}{*}{
} & $e_c H^\dagger L s$ & $s{{H^\dagger}^{i}}({{e_c}_{p}}{{L_{r}}_{i}})$ &  &  & $\xcheckmark$ &  &  \\ \cline{2-8}
\multirow{1}{*}{$\psi^2 \phi^2$} & $d_c H^\dagger Q s$ & $s{{H^\dagger}^{i}}({{{d_c}_{p}}^{a}}{{Q_{r}}_{ai}})$ &  &  & $\xcheckmark$ &  &  \\
\end{longtable}
}

\let\clearpage\relax
\section{Fermion EFT (\chiEFT) Operator Bases\label{sec:operator4fermion}}
In this section, we move to the extension with one single spin-$1/2$ fermion singlet (\chiEFT). Though we focus only on one singlet, we will consider a more general case with $n_\chi$ flavors of the singlet, as one note that a spin-$1/2$ fermion is either Majorana-type or Dirac-type which can be identified as two two-component spinors in our notation imposing further symmetries. 


\label{hyref:majorana}With one two-component fermion (Majorona-type), the most general renormalizable Lagrangian is written as
\begin{align}
    \mathcal{L}\supset&\overline{\chi}i\slashed{\partial}\chi-\left(\frac{1}{2}m_\chi\overline{\chi}\chi^c+Y\bar{\ell}_L\widetilde{H}\chi+\text{h.c.}\right),
\end{align}
where $\chi^c$ is the charge conjugation of $\chi$ and the Majorana mass terms naturally exists since the fermion is neutral under the SM gauge groups. One can identify $\chi$ as a right-handed neutrino and it helps to generate masses for neutrinos, which explains the observation of neutrino oscillations~\cite{Bahcall:1976zz, Super-Kamiokande:1998kpq, SNO:2001kpb, Pontecorvo:1957cp, Pontecorvo:1967fh}\footnote{To explain all the neutrnio oscillations observed experiments, at least two right-handed neutrinos should be introduced indeed. We only introduce one right-handed neutrino here to be consistent with our single field discussion. Readers can find the operator basis for general $n_\chi$ flavor right-handed neutrinos in Ref.~\cite{Li:2021tsq} and also Tab.~\ref{tab:fEFT}.}. Generally speaking, though only Yukawa-type interactions should be added, just following the same mechanism generating masses for quarks and charged leptons, the quantum numbers, except lepton number which is known as conserved quantity from an accidental symmetry of the SM, of the fermionic singlet do not forbid a Majorana mass term. Such Lagrangian is also known as Type-I seesaw model~\cite{Minkowski:1977sc, Yanagida:1979as, Gell-Mann:1979vob, Mohapatra:1979ia} when the Majorana mass of field $N$ is heavy (above TeV scale). A sub-GeV or keV mass of $\chi$ can serve as a DM candidate~\cite{Dodelson:1993je, Shi:1998km, Dolgov:2000ew, Abazajian:2001vt, Abazajian:2001nj, Asaka:2005an, Asaka:2005pn, Asaka:2006nq, Drewes:2016upu, Boyarsky:2018tvu}. This fermion can also serve as the portal between the visible and dark sectors if there is a dark sector which does not interact with the SM particles but only couples to this new spin-$1/2$ fermion~\cite{Falkowski:2009yz, Lindner:2010rr, Bai:2013iqa, GonzalezMacias:2015rxl, Alexander:2016aln, Battaglieri:2017aum}, but we will not go into detail on the neutrino portal case in this study since we only focus on the one field extension. At last we emphasize that the most general fermion singlet extension of the SM is identical to the sterile neutrino extension since both models have no extra symmetries except the SM gauge symmetry $SU(3)_C\times SU(2)_L\times U(1)_Y$.

At the renormalizable level, only odd number of particles $N$ (only one) is allowed in the interactions, while terms with even number of particles exist. There are only one independent operators at the next leading order (dim-5), which are given as
\begin{align}
    \mathcal{L}\supset&\frac{c_1}{\Lambda}H^\dagger H\overline{\chi}\chi^c 
    \label{eq:MF_dim5}
\end{align}
Note that Refs.~\cite{delAguila:2008ir, Aparici:2009fh, Li:2021tsq} also present a further operator $\overline{\chi}\sigma^{\mu\nu}\chi^c B_{\mu\nu}$ at this order. However, their results are for general $n_\chi$ flavors of right-handed neutrinos, while in our case of one fermion singlet, this term is forbidden by the antisymmetry of the tensor-fermion bilinear. In principle, we can impose a $\mathbb{Z}_2$ symmetry on this field to forbid the $dim$-4 Yukawa interaction and leave the leading order interaction Eq.~\ref{eq:MF_dim5}. This symmetry guarantees the stability of the fermion, making it a candidate of DM. Such model describes the Higgs-portal with spin-$1/2$ DM~\cite{Kanemura:2010sh, Lopez-Honorez:2012tov, Arcadi:2017kky}.

\label{hyref:dirac}As stated at the beginning, one can add one more two-component fermion to combine them into one Dirac fermion if and only if a global $SO(2)$ symmetry exists between them~\cite{Li:2021tsq}. 
\begin{align}
    \mathcal{L}\supset-m\chi_1\chi_1-m\chi_2\chi_2+h.c.\qquad
    \begin{pmatrix}
    \chi_1 \\
    \chi_2
    \end{pmatrix}\xrightarrow{SO(2)}\begin{pmatrix}
    \cos\phi & -\sin\phi \\
    \sin\phi & \cos\phi
    \end{pmatrix}\begin{pmatrix}
    \chi_1 \\
    \chi_2
    \end{pmatrix}.
\end{align}
We can have the left-handed fermion and a right-handed one be their complexification
\begin{align}
    \chi_L=\chi_1+i\chi_2\quad\chi_R=\chi_1-i\chi_2\longrightarrow\mathcal{L}\supset-m\overline{\chi}_L\chi_R+h.c..
\end{align}
The $SO(2)$ symmetry is isomorphic to $U(1)$ symmetry, which contributes opposite phase shifts for $\chi_L\rightarrow e^{i\phi}\chi_L$ and $\chi_R\rightarrow e^{-i\phi}\chi_R$. Therefore one can impose an exact $U(1)$ symmetry on the two two-component fermions to write down the EFT for a Dirac fermion.

The $U(1)$ symmetry forbids any renormalizable terms except the kinetic and mass terms, therefore the leading non-trivial terms shows up at the $dim$-5. In left-handed and the right-handed component notation, the leading Lagrangian reads
\begin{align}
    \mathcal{L}\supset&\overline{\chi}_L i \slashed{\partial}\chi_L+\overline{\chi}_R i \slashed{\partial}\chi_R+m\left(\overline{\chi}_L\chi_R+\text{h.c.}\right) \nonumber \\
    &+\left(\frac{c_1}{\Lambda}H^\dagger H\overline{\chi}_L\chi_R+\text{h.c.}\right)+\left(i\frac{c_2}{\Lambda}B_L^{\mu\nu}\overline{\chi}_L\sigma_{\mu\nu}\chi_R+\text{h.c.}\right)
\end{align}
where the first term in the second line parameterizes the interaction between a Dirac fermionic DM and the SM via the Higgs portal $H^\dagger H$~\cite{Kim:2006af, Kim:2008pp, Fedderke:2014wda}, and the second term can induce the electromagnetic dipole interaction, which might lead to interesting signatures in both DM detection and collider experiments~\cite{Sigurdson:2004zp, Banks:2010eh, Fortin:2011hv, Weiner:2012cb}.

In the following Tab.~\ref{tab:fEFT}, we present the complete non-redundant operator basis up to $dim$-8 for $n_\chi$ Majorana fermions extension of the SM in two-component fermion notation. Both single Majorana fermion extension (\mchiEFT) and Dirac fermion extension (\dchiEFT) can be obtained by applying further constraints: 1) setting $n_\chi=1$, one arrives at single Majorana fermion case (\textbf{M}); and 2) setting $n_\chi=2$ and assigning two fermions with opposite $U(1)$ charges, one arrives at single Dirac fermion case (\textbf{Dirac}).

\setlength\LTleft{-0.7in}
{ \small 
\begin{longtable}{@{}C{0.09\textwidth}C{0.13\textwidth}C{0.375\textwidth}C{0.02\textwidth}C{0.315\textwidth}C{0.09\textwidth}C{0.03\textwidth}@{}}
\caption{\textcolor{black}{{Complete and independent operators for $n_\chi$ Majorana fermions (\textbf{$n_\chi$ Majorana}), single Majorana (\textbf{M}), Dirac fermion (\textbf{Dirac}), $\mathbf{Z}_2$ dark matter ($\mathbf{Z}_2$) classified in canonical dimension. \label{tab:fEFT}}}}\\
\hhline{=======}
\multicolumn{7}{c}{\textcolor{blue}{{\textbf{\textit{dim}-4 operators}}}}\\
\hhline{=======}
{\textbf{Class}} & {\textbf{Type}} & {\textbf{$n_\chi$ Majorana}} & {\textbf{M}} & {\textbf{Dirac}} & {\textbf{F}} & {$\mathbf{Z}_2$} \\
\hline\hline
$\phi\psi^2$ & $HL\chi_c$ & ${\epsilon^{ij}}{H_{j}}({{L_{p}}_{i}}{{\chi_c}_{r}})$ &  &  &  &  \\ \hline

\hhline{=======}
\multicolumn{7}{c}{\textcolor{blue}{{\textbf{\textit{dim}-5 operators}}}}\\
\hhline{=======}
{\textbf{Class}} & {\textbf{Type}} & {\textbf{$n_\chi$ Majorana}} & {\textbf{M}} & {\textbf{Dirac}} & {\textbf{F}} & {$\mathbf{Z}_2$} \\
\hline\hline
$F_L\psi^2$ & $B_L \chi_c^2$ & ${{B_L}^{\mu\nu}}({{\chi_c}_{p}}{\sigma_{\mu\nu}}{{\chi_c}_{r}})$ & \xmark & ${{B_L}^{\mu\nu}}({{\chi_L}_c}{\sigma_{\mu\nu}}{{\chi_R}_c})$ &  & $\checkmark$ \\ \hline

$\phi^2\psi^2$ & $HH^\dagger \chi_c^2$ & ${H_{i}}{{H^\dagger}^{i}}({{\chi_c}_{p}}{{\chi_c}_{r}})$ &  & ${H_{i}}{{H^\dagger}^{i}}({{\chi_L}_c}{{\chi_R}_c})$ &  & $\checkmark$ \\
\end{longtable}
}

\let\clearpage\relax
\section{Vector EFT (\VEFT) Operator  Bases\label{sec:operator4photon}}
Unlike the scalar and fermion cases, to have spin-1 vector fields generally requires to extend the gauge structure of the SM if one wants to keep a local description. The simplest and most interesting case is introducing an extra $U(1)$ symmetry, whose gauge boson is named as dark photon~\cite{Okun:1982xi, Galison:1983pa, Holdom:1985ag}~\footnote{The names para-, hidden-sector and secluded photon are also being used to indicate the same particle.}. Though dark photon can be either massless and massive, we stick to the massive one in the discussion since the massless case can easily obtained by ignoring the operators with four-vector potential field ($X_\mu$)~\footnote{The massless vector field should correspond to some non-broken gauge symmetry. And to respect such gauge symmetry, all the operators should be built from field strength tensor not four-vector potential field in our method.} and usually receives more severe constraints. Such a massive vector can account for part or whole of dark matter if it is stable~\cite{Nelson:2011sf, Arias:2012az, Graham:2015rva}, or it provides us the third kind portal (vector portal) if dark matter is charged under the $U(1)$ symmetry~\cite{Knapen:2017xzo, Hambye:2019dwd, Fabbrichesi:2020wbt, Alexander:2016aln, Battaglieri:2017aum}.

At the lowest order, the interaction between the real massive vector field (also known as dark photon) and the SM can be describe by Proca Lagrangian~\cite{Proca:1936fbw} with an addition kinetic mixing term to hypercharge
\begin{align}
    \mathcal{L}\supset-\frac{1}{4}X_{\mu\nu}X^{\mu\nu}+\frac{1}{2}m_X^2 X_\mu X^\nu-\epsilon B_{\mu\nu}X^{\mu\nu},
\end{align}
where $X_\mu$ and $B_\mu$ are extra $U(1)$ and hypercharge ($U(1)_Y$) vector fields in gauge basis respectively. Such kinetic mixing is always possible since the field strengths of two Abelian gauge fields can be multiplied together to give an invariant $dim$-4 operator.

It is well-known that a mass term for vector field breaks the gauge symmetry explicitly and UV completion is needed for the Proca Lagrangian. From the EFT point of view, one can directly write down the operators with a vector fields by requiring only Lorentz symmetry but ignoring its corresponding gauge symmetry, like the one obtained in Ref.~\cite{Criado:2021trs}. However, one should note that the only consistent spin-1 field must be a gauge field. The previous mentioned procedure buries the fundamental gauge structures, which are usually helpful to organize operators with further constraints. Therefore, some operators forbidden by the fundamental gauge symmetries~\cite{Arzt:1994gp, Bhattacharya:2021edh} are included in Ref.~\cite{Criado:2021trs}, which might result in problems to UV complete the effective theories. However, if one insists on gauge symmetry, the effective operators can be only built from field strength tensors. Ref.~\cite{Aebischer:2022wnl} then only adds mass terms for vector fields, which only generates an incomplete basis. So a consistent gauge invariant building method benefits the operators reducing procedure and is more convenient to find the non-redundant basis. 

The only known mechanism to let vector fields obtain their mass is through the spontaneous breakdown of a local symmetry, which is generally described by the Higgs mechanism~\cite{Higgs:1964pj, Englert:1964et}. However, for $U(1)$ gauge symmetry, another well known mechanism to preserve gauge symmetry while generating mass for vector boson is to introduce pure gauge compensating (Stueckelberg) field~\cite{Stueckelberg:1938hvi, Ruegg:2003ps}. In our previous paper~\cite{Song:2023lxf}, we emphasize again that the Stueckelberg mechanism can be identified as a special case of the Higgs mechanism where the Higgs particle is decoupled (which is also known as affine limit) and Stueckelberg field is just a Goldstone boson. Therefore we build a complete set of operators for a dark photon by introducing an extra Goldstone field and then choosing a certain gauge~\footnote{Such kind of idea is also discussed in Ref.~\cite{Lebedev:2011iq, Duch:2014xda, Bhattacharya:2021edh}.}. Such auxiliary Goldsone not only helps to recover the corresponding gauge symmetry but also is phenomenologically useful for studies of high-energy behaviour and UV completion.

Similar to scalar and fermion cases, one can double the DOFs and combine the two real vectors to form a complex vector. Such complex vector singlet extension is less discussed, since there is no renormalizable operator except the kinetic term due to the factor that the vector carries a non-trivial charge of an extra $U(1)$ symmetry. However, it commonly exists, though usually with some extra states, if some hidden local $SU(2)$ symmetry spontaneously breaks into a $U(1)$ symmetry, like in Twin Higgs model~\cite{Chacko:2005pe}. From EFT's perspective, we also consider such complex vector case not just for completeness but believe that it must be phenomenological interesting.

In the following Tab.~\ref{tab:VEFT}, we present complete and independent operators of \VEFT~for either a real or complex massive vector.

\setlength\LTleft{-0.6in}
{ \small 
\begin{longtable}{@{}C{0.09\textwidth}C{0.12\textwidth}C{0.36\textwidth}C{0.365\textwidth}C{0.1\textwidth}c@{}}
\caption{\textcolor{black}{{Complete and independent operators for dark photon written in Stueckelberg mechanism with Goldstone (\textbf{Real}), unitary gauge with massive vector field (\textbf{Complex}), $\mathbf{Z}_2$ dark matter ($\mathbf{Z}_2$) classified in canonical dimension. The operators for a massless dark photon can be read directly from \textbf{Complex} column by keeping only the operators with $\checkmark$ marks.\label{tab:VEFT}}}}\\
\hhline{======}
\multicolumn{6}{c}{\textcolor{blue}{{\textbf{\textit{dim}-4 operators}}}}\\
\hhline{======}
{\textbf{Class}} & {\textbf{Type}} & {\textbf{Real}} & {\textbf{Complex}} & {\textbf{F}} & {$\mathbf{Z_2}$}  \\
\hline\hline
$F_L^2$ & $X_L B_L$ & ${B_L}_{\mu\nu}{X_L}^{\mu\nu}$ & 
 &  \\ \hline

\multirow{5}{*}{
} & $u e_c e_c^\dagger$ & $({{e_c}_{p}}{\sigma^{\mu}}{{e^\dagger_c}_{r}})X_\mu$ & \textcolor{red}{
} &  \\ \cline{2-6}
 & $u LL^\dagger$ & $({{L_{p}}_{i}}{\sigma^{\mu}}{{{L^\dagger}_{r}}^{i}})X_\mu$ & \textcolor{red}{
 } &  \\ \cline{2-6}
\multirow{1}{*}{$u\psi\bar{\psi}$} & $u QQ^\dagger$ & $({{Q_{p}}_{ai}}{\sigma^{\mu}}{{{Q^\dagger}_{r}}^{ai}})X_\mu$ & \textcolor{red}{
} &  \\ \cline{2-6}
 & $u d_c d_c^\dagger$ & $({{{d_c}_{p}}^{a}}{\sigma^{\mu}}{{{d^\dagger_c}_{r}}_{a}})X_\mu$ & \textcolor{red}{
 } &  \\
\end{longtable}
}

\let\clearpage\relax
\section{Conclusion \label{sec:conclu}}

\begin{table}[h]
\makebox[1 \textwidth][c]{       
\resizebox{1 \textwidth}{!}{   
\begin{tabular}{c|c|c|c|c|c|c|c}
\hline\hline
\multicolumn{3}{c|}{Singlet} & $dim$-4 & $dim$-5 & $dim$-6 & $dim$-7 & $dim$-8 \\ \hline\hline

\multirow{6}{*}{\hyperref[sec:operator4scalar]{Scalar}} & \multirow{3}{*}{\hyperref[hyref:real]{Real}} & w/o $\mathbf{Z_2}$ & - & $9+6n_f^2$ & $10+n_f+7n_f^2$ & \thead{$30+n_f+\frac{965}{12}n_f^2$\\$+\frac{3}{2}n_f^3+\frac{397}{12}n_f^4$} & \thead{$\frac{1}{12}(516+36n_f+1241n_f^2$\\$+42n_f^3+661n_f^4)$} \\ \cline{3-8}
& & w/ $\mathbf{Z_2}$ & - & - & $10+6n_f^2$ & $n_f+n_f^2$ & \thead{$\frac{1}{12}(516+1085n_f^2$\\$+18n_f^3+397n_f^4)$} \\ \cline{2-8}
& \multicolumn{2}{c|}{\multirow{1}{*}{\hyperref[hyref:complx]{Complex}}} & - & - & $12+11n_f^2$ & $n_f+n_f^2$ & \thead{$58+\frac{1745}{12}n_f^2$\\$+\frac{3}{2}n_f^3+\frac{397}{12}n_f^4$}\\ \hline\hline

\multirow{3}{*}{\hyperref[sec:operator4fermion]{Fermion}} & \multirow{2}{*}{\hyperref[hyref:majorana]{Majorana}} & w/o $\mathbf{Z_2}$ & $2n_f$ & $2$ & \thead{$2+8n_f$\\$+6n_f^2+13n_f^3$} & \thead{$\frac{4}{3}(12+14n_f$\\$+9n_f^2+25n_f^3)$} & \thead{$18+68n_f$\\$+56n_f^2+181n_f^3$} \\ \cline{3-8}
& & w $\mathbf{Z_2}$ & - & $2$ & $2+5n_f^2$ & $4(4+3n_f^2)$ & $18+2n_f+57n_f^2$\\ \cline{2-8}
& \multicolumn{2}{c|}{\multirow{1}{*}{\hyperref[hyref:dirac]{Dirac}}} & - & $4$ & $7+10n_f^2$ & $22+28n_f^2$ & $43+n_f+113n_f^2$ \\ \hline\hline

\multirow{6}{*}{\hyperref[sec:operator4photon]{Vector}} & \multirow{3}{*}{\hyperref[hyref:real]{Real}} & w/o $\mathbf{Z_2}$ & $4+5n_f^2$ & - & $37+71n_f^2$ & $\frac{2}{3}(8n_f^2+7n_f^4)$ & \thead{$\frac{1}{12}(4836+13529n_f^2$\\$+6n_f^3+4477n_f^4)$} \\ \cline{3-8}
& & w/ $\mathbf{Z_2}$ & 2 & - & $22+21n_f^2$ & $n_f+n_f^2$ & \thead{$\frac{1}{6}(1152+2503n_f^2$\\$+231n_f^4)$} \\ \cline{2-8}
& \multicolumn{2}{c|}{\multirow{1}{*}{\hyperref[hyref:complx]{Complex}}} & 3 & - & $51+42n_f^2$ & $n_f^2$ & \thead{$\frac{1}{3}(1617+2579n_f^2$\\$+3n_f^2+346n_f^4$} \\ \hline\hline
\end{tabular} }}
\caption{Summary of the complete sets of operator basis for single singlet particle extensions of the SMEFT (\sEFT), depending on the natures of the singlet, like spin, hermiticity, $\mathbf{Z}_2$ symmetry, for $n_f$ flavor generations.~\label{tab:OBssum}}
\end{table}

In this work, we present the complete and independent operators of WILP extensions of the SM, including real/complex scalar, Majorana/Dirac fermion, and real/complex vector, up to dimension eight, by means of the on-shell amplitude method with Young tensor technique. The EFTs for complex or Dirac particles can be easily obtained by double the corresponding real or Majorana DOFs with adding an extra $U(1)$ symmetry. Imposing further a discrete $\mathbf{Z}_2$ on the singlets, these particles can be stable and work as WIMP dark matter candidates. For the scalar case, the scalar itself can be non-fundamental but a (pseudo-)Nambu-Goldstone remaining after spontaneous breaking of some global symmetry. Axion, ALPs and Majoron fall into this category. The shift symmetry, or the derivative coupling, of the Goldstone nature can be expressed into the Adler's zero condition in the language of S-matrix. The complete non-redundant bases for them are further reduced to subspaces of the scalar basis. General operators of massive vector extensions of the SM can be built by combining Goldstone operator basis and massless $U(1)$ gauge basis. The complete sets of operator basis, investigated in this work, are summarized in Tab.~\ref{tab:OBssum}. All the numbers of operators are further cross-checked with a Hilbert series counting, which shows consistency.

Such complete operator bases would benefit various phenomenological studies. On one hand, the complete bases are necessarily needed during matching between the UV physics and the effective theories; on the other hand, the experimental signals coming from the higher dimensional operators would be comparable to the ones from well-studied leading operators, especially in UV models where the leading operators are suppressed either by some symmetries or by accidental cancellations. Further, based on power counting rules, one would expect that the effects from next-to-leading operators should be comparable to the effects calculated at the one-loop level with leading operators, which have been considered in recent years. We hope that our bases would benefit the experimental searches for dark matter and light particles.


	
\section*{Acknowledgments} 
This work is supported by the National Science Foundation of China under Grants No. 12022514, No. 11875003 and No. 12047503, and National Key Research and Development Program of China Grant No. 2020YFC2201501, No. 2021YFA0718304, and CAS Project for Young Scientists in Basic Research YSBR-006, the Key Research Program of the CAS Grant No. XDPB15. 



%


\bibliographystyle{JHEP}
\bibliography{ref}

\end{document}